\documentclass[aps,
twocolumn,
showpacs,
prb,
amsmath,amssymb,floatfix,superscriptaddress]{revtex4-2}
\usepackage{CJKutf8}
\usepackage{amsmath,amssymb,graphicx,color}
\usepackage{bm}
\usepackage{bbm}
\usepackage[caption=false]{subfig}  
\usepackage[usenames,dvipsnames]{xcolor}
\usepackage{mathrsfs}
\usepackage[mathscr]{euscript}

\usepackage[normalem]{ulem}

\usepackage{soul}
\usepackage[colorlinks=true,citecolor=Cerulean,linkcolor=RubineRed,urlcolor=Cerulean]{hyperref}
\usepackage{soul,xcolor}
\usepackage{braket}
\usepackage{makecell}

\usepackage{cleveref}

\newcommand{\unit}[1]{\,\mathrm{#1}} % for specifying units in math mode
\newcommand{\equa}[1]{Eq.~\eqref{#1}} 
\newcommand{\fig}[1]{Fig.~\ref{#1}}

\usepackage{ulem}

\newcommand{\rom}[1]{\uppercase\expandafter{\romannumeral #1\relax}}

\usepackage{ulem}
\usepackage[T1]{fontenc}
\usepackage{times}{\Large}

\begin{document}
\title{Interaction effects on electronic Floquet spectra: Excitonic effects}

\author{Teng Xiao}
\affiliation{State Key Laboratory of Low-Dimensional Quantum Physics and Department of Physics, Tsinghua
	University, Beijing 100084, People's Republic of China}

\author{Tsan Huang}
\affiliation{State Key Laboratory of Low-Dimensional Quantum Physics and Department of Physics, Tsinghua
	University, Beijing 100084, People's Republic of China}
	
\author{Changhua Bao}
\affiliation{State Key Laboratory of Low-Dimensional Quantum Physics and Department of Physics, Tsinghua
	University, Beijing 100084, People's Republic of China}

\author{Zhiyuan Sun} \email{Contact author: zysun@tsinghua.edu.cn}
\affiliation{State Key Laboratory of Low-Dimensional Quantum Physics and Department of Physics, Tsinghua
University, Beijing 100084, People's Republic of China}
\affiliation{Frontier Science Center for Quantum Information, Beijing 100084, People's Republic of China}

\date{\today}
\begin{abstract}
Floquet engineering of electronic states by light is a central topic in modern experiments. However, the impact of many-body interactions on the single-electron properties remains unclear in this non-equilibrium situation. 
We propose that interaction effects could be reasonably understood by performing perturbative expansion in both the pump field and the electron-electron interaction when computing physical quantities.
As an example, we apply this approach to semiconductors and show analytically that excitonic effects, i.e., effects of electron-hole interaction, lead to dramatic corrections to the single-electron Floquet spectra even when the excitons are only virtually excited by the pump light. We compute these effects in phosphorene and monolayer MoS$_2$  for  time- and angle-resolved photoemission spectroscopy and ultrafast optical experiments.	
\end{abstract}

\maketitle

%{\emph{Introduction}---}
As a central arsenal of modern experiments, ultrafast lasers have  been widely used to manipulate and explore the electronic properties in solid-state materials.
When the laser pulse (the pump) is a multi-cycle one, the pumped material could be treated as a periodically driven system, giving rise to non-equilibrium states widely described in the language of  Floquet engineering~\cite{Oka:2019aa,Rudner:2020aa,Kennes.2021_rmp,Zhou.2022_review,Oka2009prb}.
For example, the time- and angle-resolved photoemission spectroscopy (Tr-ARPES) could generate the Floquet electronic  states and directly probe their  energy-momentum band dispersion~\cite{Andrea2024review,Wang2013Science,Mahmood:2016aa,Aeschlimann2021,Zhou_Nature_2023, Ito2023,Zhou_PRL_2023,EcksteinPRB2008}.
On the other hand, the impact of many-body interactions on the Floquet electronic spectra is still a complicated problem that is being actively studied~\cite{Bukov2015prl,Mikhaylovskiy2015,Kennes2018,Claassen:2017_spin_liquid_floquet,Wang2018prl,DasariPRB2018,LiuPRB2019,YatesPRB2019,LiPRB2020,BittnerPRB2020,YatesPRB2021,LiPRB2021,LenkPRB2022,QuitoPRB2023,xie2023arxiv}.
The conventional theoretical approach computes the exact single-electron Floquet bands as the first step, which typically has to be done numerically~\cite{Sentef2015,SCHULER2020107484, Liu_2023, Cao2024}, and then performs perturbation in the electron-electron  interaction on top of it. This method is limited by substantial numerical complexity and the  lack  of analytical clarity.

In this Letter, we show that interaction effects in electronic Floquet bands could be understood analytically from a simpler  approach, which performs perturbative expansion in both the pump field~\cite{Sun_Floquet_PRB_2024} and the electron-electron interaction.
By deriving  the dramatic excitonic and phononic corrections to the Floquet electronic bands in semiconductors, we show that this method enjoys both  physical clarity and computational efficiency.

\begin{figure}[tbp]
	\centering \includegraphics[width=1\linewidth]{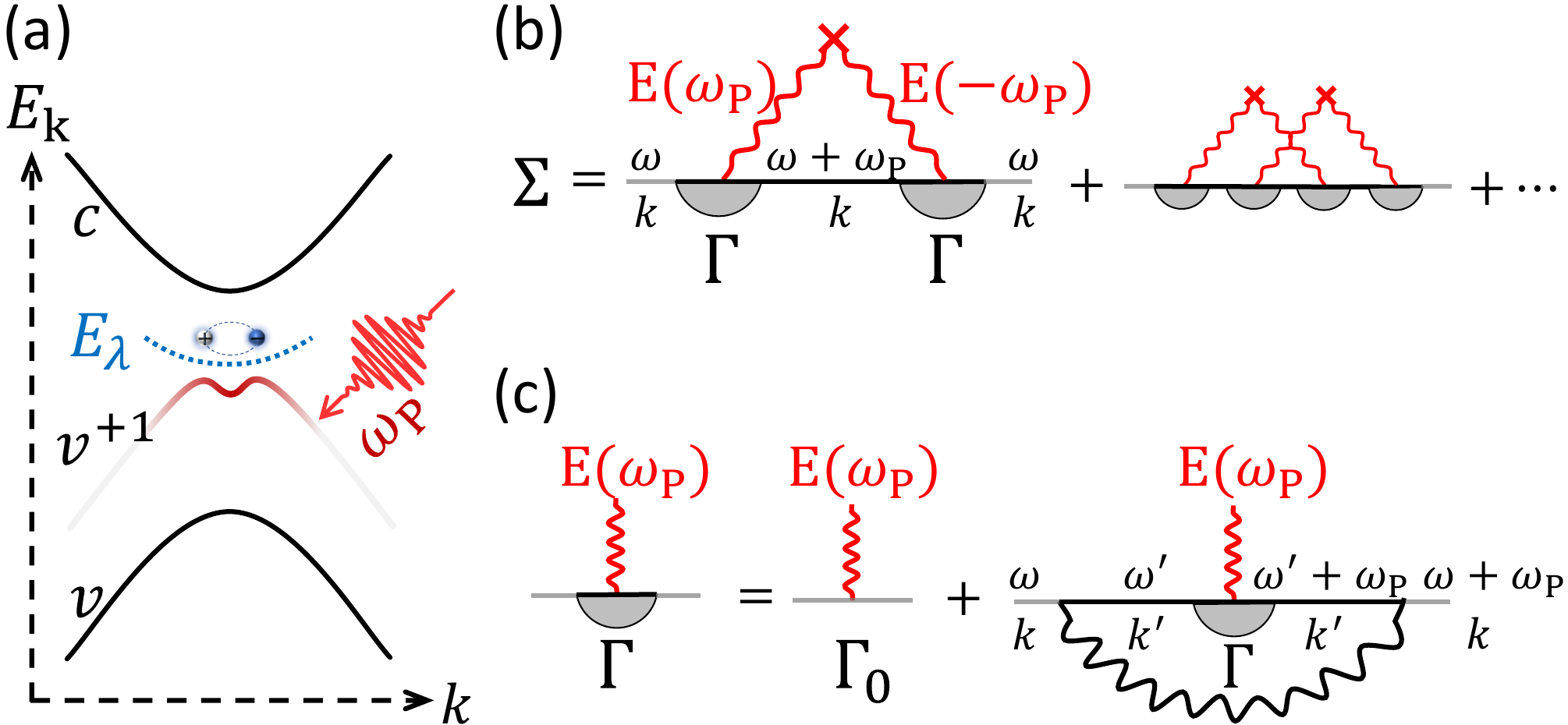} \caption{(a) Schematic of a semiconductor with conduction band $c$ and valence band $v$ represented by black lines. Driven by pump light at frequency $\omega_{\text{P}}$, the electron is virtually pumped to a  Floquet side band $v^{+1}$ (red curve), which is dramatically corrected by the virtual excitation of the zero-momentum excitonic mode on the blue dashed curve. 
		(b), (c) The Feynman diagrams of the self-energy $\Sigma$ and the light-matter vertex $\Gamma$. 	The red wavy lines represent the electric field of the pump light connected in pairs to the ``disorder'' vertices. The black wavy lines are the electron-hole interaction. 
	}
	\label{fig1}
\end{figure}

 For example, when computing the electronic Green function (propagator)
\begin{equation}\label{eqn:dyson}
	\begin{aligned}
		{G} & =\left(\begin{array}{cc}
			G^R & G^K \\
			0 & G^A
		\end{array}\right)
		=\left({G}_0^{-1}-{\Sigma}\right)^{-1}
	\end{aligned}
\end{equation}
from the  equilibrium bare Green function ${G}_0$,
the self-energy ${\Sigma}$  could be  perturbatively expanded in the driving field $E(t)=\int d\omega_{\text{P}}  E(\omega_{\text{P}}) \mathrm{e}^{-\mathrm{i}\omega_{\text{P}} t}$, as shown by the red wavy lines in \fig{fig1}(b). 
In non-equilibrium systems, instead of imaginary-time Green functions, calculations should be performed in principle  using real-time Green functions directly~\cite{Kadanoff.1962,Keldysh.1964, Kamenev.book, Altland.2010, Sieberer.2016}, for which the neatest notation is the two-by-two Green function in \equa{eqn:dyson}  that includes the retarded ($G^R$), advanced ($G^A$), and Keldysh ($G^K$) components introduced by Keldysh~\cite{Keldysh.1964, Kamenev.book, Altland.2010, Sieberer.2016}; see the Supplemental Material (SM), Sec.~I, for definitions~\cite{Supp}. 
This Green function contains the information of nearly all single-particle properties such as the energy dispersion, occupation and spectral weight, and is the central quantity in this paper.

The Green function $G(t_0+t/2,t_0-t/2)$ could be written  in the frequency representation as $G(\omega, t_0)=\int_{-\infty}^\infty dt e^{i\omega t}G(t_0+t/2,t_0-t/2)$.
For the nonequilibrium system whose Hamiltonian oscillates and loses time-translational invariance, $G(\omega, t_0)$ depends on the central time $t_0$.
Since  its $t_0$ dependence is typically in a fast oscillating manner during the pump pulse which can hardly be resolved in ultrafast experiments, 
we focus on its time-averaged component
$G(\omega)=\int_0^T G(\omega, t_0) dt_0/T$, where $T$ is a timescale that is much larger than the typical period of the pump light. It reduces to the equilibrium Green function in the limit of zero driving field. 
If one puts it in the context of the Green function $G_{m,n}(\omega)$ of  a strictly periodically driven system where $m,n$ are integers~\cite{Aoki2014}, the Green function used here is  $G(\omega)=G_{0,0}(\omega)$. However, note that our approach is not limited to periodically driven systems and could also deal with the case of an ultrafast pump, i.e., when the frequency spectra $E(\omega_{\text{P}})$ is not a $\delta$ function.

For the Green function of interest, the perturbative expansion  of the self-energy $\Sigma$ in the pump field is similar to that for the disorder potential; see Fig.\,\ref{fig1}(b). 
It contains every diagram that satisfies two conditions: (1) it is one-particle-irreducible (1PI), meaning that it cannot be separated by cutting a electron propagator with the frequency labeling $\omega$; and (2) there are equal numbers of lines of the driving field $E$ at the frequencies $\omega_{\text{P}}$ and $-\omega_{\text{P}}$  so that the self-energy does not change the frequency.
Every pair of driving field lines connected to the same ``disorder'' vertex gives a Gaussian-ensemble-averaged driving power, $\langle E(\omega_{\text{P}}) E(\omega_{\text{P}}^\prime) \rangle = S(\omega_{\text{P}})\delta(\omega_{\text{P}}-\omega_{\text{P}}^\prime)$, which applies to the case when the pump pulse has random phases.
The lowest-order self-energy is therefore
\begin{equation}\label{eqn:self_energy}
	\begin{aligned}
		\Sigma  =& \int d\omega_{\text{P}}  \Gamma(-\omega_{\text{P}})G_{0}(\omega+\omega_{\text{P}}) \Gamma(\omega_{\text{P}})
    \end{aligned}
\end{equation}
as shown by the first diagram in \fig{fig1}(b) where $\Gamma$ is the electron-photon vertex  shown in \fig{fig1}(c). The Dyson's equation (\equa{eqn:dyson})  yields a Green function  to infinite order in $\Sigma$. 
Perturbation in electron-electron, electron-phonon or electron-impurity interactions could be added diagrammatically in similar ways to the equilibrium situation, e.g., as corrections to the  vertex $\Gamma$ or to the bare Green function $G_0$.  
For materials driven by light, the most dramatic effect occurs when an optically active collective mode, such as an infrared phonon or exciton that couples to the electrons~\cite{Hubener2018,BiontaPRR2021,Freudenstein2022, OferPNAS2022,ZhangNP2024,ZhangnpjCM2024,Kobayashi2023,ChangLee2024}, is  driven linearly  by light. This effect is contained in the  interaction correction to the electron-photon vertex $\Gamma$, as shown by \fig{fig1}(c) for the case of excitonic correction, i.e., correction from electron-hole interaction.

\emph{Excitonic corrections.} We now apply this method to compute the excitonic correction to the Floquet electronic bands in semiconductors driven by  light. We show analytically that the optically active excitonic modes lead to dramatic corrections to  both the energy and occupation of the bands. Predictions are shown for monolayer black phosphorus (phosphorene) and monolayer MoS$_2$ to be verified by Tr-ARPES or ultrafast optical experiments.
We note that although people have tried to calculate the Tr-ARPES signatures of excitons with numerical methods~\cite{Perfetto_PRB_2016,Rustagi_PRB_2018,Christiansen_PRB_2019, Perfetto.2019, Chan_Giant_PNAS_2023,Park_2024,Schuler.2024,ParkPRB2025}, this work provides a unified analytical method  that offers physical understanding.

A minimal two band Hamiltonian for  semiconductors with electron-hole interaction and nonzero interband optical matrix element is 
\begin{align}\label{eq:model}
		&{H}=\sum_{{k}} 
		\left(
		\begin{array}{c}
			{c}_{k}	\\
			{v}_{k}\\
		\end{array}
		\right)^\dagger \left(
		\begin{array}{cc}
			\varepsilon_{k+A}^c&A \cdot M_k\\
			A \cdot M^{*}_k&\varepsilon_{k+A}^v\\
		\end{array}
		\right) \left(
		\begin{array}{c}
		{c}_{k}	\\
			{v}_{k}\\
		\end{array}
		\right)  
		+H_{\text{ee}},
\notag\\
	 &H_{\text{ee}}= \sum_{k, k^{\prime}, q} V_q  {v}_{k^{\prime}-q}^{\dagger} {v}_{k^{\prime}} {c}_{k+q}^{\dagger} {c}_{k}
\end{align}
where $\varepsilon_{k}^v=-{ k^2}/{2 m_v}$ ($\varepsilon_{k}^c=E_{\text{g}}+{ k^2}/{2 m_c}$) is the kinetic energy of the valence-band (conduction-band) electrons whose  annihilation operators are $v_k$ ($c_k$) at momentum $k$, $E_{\text{g}}>0$ is the band gap, and $V_q$ is the density-density interaction kernel between electrons of the two bands.   For notational simplicity, we set the Planck constant $\hbar$, the elementary charge $e$ and the speed of light $c$ to be 1.

For clear illustration of the physics, we use the single-color pump light with the electric field $E=-\partial_t A$ represented by the vector potential    ${A}(t)={A}_0\mathrm{e}^{-\mathrm{i}\omega_{\text{P}} t}+\text{c.c.}$, so that one does not need to perform the integral in $\omega_{\text{P}}$ in \equa{eqn:self_energy}.
The vector potential couples to electrons through the Peierls substitution and the  interband optical matrix element
$A_0 M_k = -\mathrm{i}E_0 M_k/\omega_{\text{P}}$, so that \equa{eq:model} is not manifestly gauge invariant.

From single-particle analysis neglecting $H_{\text{ee}}$, as the driving field is turned on, the electronic bands  become Floquet bands (contained in the poles of the Green function), such as the $v^{+1}$ band shown in Fig. \ref{fig1}(a). Taking into account the electron-hole attraction $H_{\text{ee}}$, there are electron-hole bound-state excitations within the gap in the excitation spectrum: the bosonic excitations called excitons.
Since the pump light couples to the interband transition terms through $M_k$, it could linearly drive the excitonic excitations in the same way as driving harmonic oscillators. More interestingly, when the pump frequency $\omega_{\text{P}}$ is close to an exciton resonance, the forced excitonic oscillation is resonantly enhanced, so that one expects dramatic interaction corrections to the electronic Floquet bands.

We now calculate this excitonic correction analytically, which is contained in the interaction modified vertex $\Gamma_k$ shown in Fig.\,\ref{fig1}(c).
Summing the ladder diagrams renders
\begin{equation}\label{eqn:vertex}
\Gamma^{cv}_k(\omega_{\text{P}})
=A_0\sum_{\lambda}
C_\lambda
\frac{\omega_{\text{P}}+\varepsilon^v_k-\varepsilon^c_k+\mathrm{i}0^+}{\omega_{\text{P}}-\omega_\lambda+\mathrm{i}\gamma_{\lambda}}
\phi_\lambda(k)
\end{equation}
where $\lambda$ labels the electron-hole two-body eigen state $\phi_\lambda(k)$ with the eigen energy  $\omega_\lambda$ satisfying the Wannier equation
\begin{equation}\label{eq:wannier}
	\left(\varepsilon^c_k-\varepsilon^v_k-\omega_\lambda\right) \phi_\lambda(k)=\sum_{k^{\prime}} V_{k-k^{\prime}} \phi_\lambda(k^{\prime})
,
\end{equation}
see the SM, Sec.~II~\cite{Supp}.
The coefficient
$C_\lambda =\sum_{k^{\prime}}M_{k^{\prime}}\phi^*_\lambda(k^{\prime})$ is the 
overlap between the interband optical matrix element and $\phi_\lambda(k)$.
The pole structure in \equa{eqn:vertex} comes from the retarded Green function of the excitons, to which we have added the phenomenological damping rate $\gamma_{\lambda}$.
Note that the index $\lambda$ contains both the bound and scattering states while only bound states with $\omega_\lambda < E_{\text{g}}$ correspond to  excitons.
Plugging \equa{eqn:vertex} (and $\Gamma^{vc}, \Gamma^{cc}, \Gamma^{vv}$) into Eqs.~\eqref{eqn:self_energy} and~\eqref{eqn:dyson} yields the electronic self-energy and Green function. The poles of the latter give the energies of the Floquet bands.
 
To make connections to the conventional single particle  approach to Floquet bands, the vertex is an energy scale that hybridizes the corresponding bare Floquet replica of the bands.
The interaction-corrected Floquet bands are the same as those from diagonalizing the Floquet Hamiltonian (with a cutoff of $\pm \omega_{\text{P}}$)  from \equa{eq:model} without $H_{\text{ee}}$, but with the interband optical matrix element $AM_k$ ($A  M^{*}_k$) replaced by the interaction-corrected vertex $\Gamma^{cv}_k$ ($\Gamma^{vc}_k$); see the SM, Sec.~V~\cite{Supp}.

\begin{figure*}[t]
	\centering \includegraphics[width=1\linewidth]{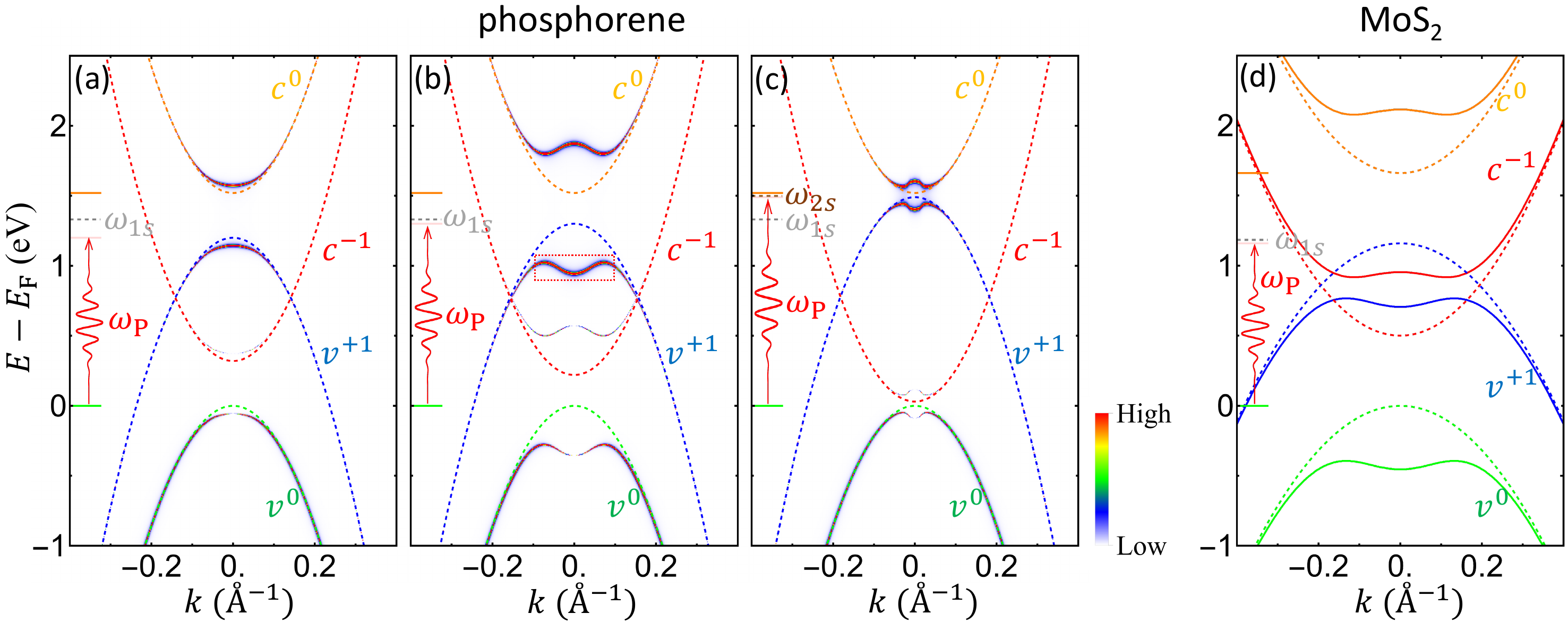} \caption{(a)-(c) Floquet electronic spectra of phosphorene  driven by pump light at frequencies $\omega_{\text{P}}=1.200 \unit{eV}$, $1.300\unit{eV}$, and $1.490\unit{eV}$, respectively.
	The colored dashed curves labeled by $c^0$, $c^{-1}$, $v^{+1}$, and $v^0$ are the four low-energy Floquet electronic  bands predicted by free theory without interaction corrections. 
		The color map shows the photoemission intensity of the interaction corrected Floquet bands.  
		The equilibrium conduction-band minimum (valence-band maximum) is indicated by the orange (green) horizontal line. 
		The gray dashed line marks the energy of the $1s$ exciton ($E_{1s}=1.331$\,eV). The brown dashed line in (c) marks the energy of the $2s$ exciton ($E_{2s}=1.499$\,eV).
		Here, $m_c=0.17m_e$, $m_v=0.18m_e$, $E_{\text{g}}=1.52 \unit{eV}$, $\epsilon=5$, $\gamma_{1s}=\gamma_{2s}=15 \unit{meV}$~\cite{Tian.2020, Yoon.2021}, and $M_k=\mathrm{i}5.25 \unit{eV \text{\AA}}$~\cite{Pereira_PRB_2015}.
		The  pump electric field  is  $E_0=5\times 10^5 \unit{V/cm}$ and along the armchair ($x$) direction of phosphorene.
		The probe matrix elements are constrained by the symmetry of  phosphorene: $M_{{fc}}=b_1$ and $M_{{fv}}=-b_1 M_k k/E_{\text{g}}$, where we set $b_1=1$, the probe incident plane is the $z-y$ plane, and the probe field is in the zigzag ($y$) direction~\cite{Bao2024,Fan2025SA}. 
		(d) Similar to (a)-(c), but for the lowest conduction band and highest valence band of monolayer $1$H-MoS$_2$ with momenta measured from the $K$ point. The Floquet bands with excitonic corrections are instead shown by solid lines.
		The parameters are $m_c=0.40 m_e$, $m_v=0.48 m_e$,   $E_{\text{g}}=1.66 \unit{eV}$, $\epsilon=5$, $E_{1s}=1.18 \unit{eV}$, $\gamma_{1s}=\gamma_{2s}=5 \unit{meV}$, $M_k=3.51 \unit{eV \text{\AA}}$, $\omega_{\text{P}}=1.16 \unit{eV}$ and $E_0= 2.9\times 10^5 \unit{V/cm}$.
		}
	\label{fig2}
\end{figure*}

The physical consequence is more transparent if the pump frequency $\omega_{\text{P}}$ is close to the eigen energy $\omega_{\lambda}$ of a certain exciton. Here the vertex will be dominated by this exciton,
\begin{equation}\label{eqn:delta_k}
\begin{aligned}
\Gamma_k^{cv}(\omega_{\text{P}})
		\approx A_0 M\frac{\omega_{\text{P}}+\varepsilon^v_k-\varepsilon^c_k}{\omega_{\text{P}}-\omega_\lambda+\mathrm{i}\gamma_{\lambda}} \phi^*_\lambda(r=0)\phi_\lambda(k)
		\equiv \Delta_k
\end{aligned}
\end{equation}
where we assumed that $M_k=M$ so that the overlap coefficient  becomes $C_\lambda=M\phi^*_\lambda(r=0)$ and that  $\Delta_k$ is proportional to $\phi_\lambda(r=0)$, the probability amplitude that the electron and hole make a contact within the bound state. Physically, the local interband optical matrix element means that light excites the exciton by generating an electron and a hole at the same space location.
From the poles of the retarded Green function $G^R$, the energy dispersion of the Floquet electronic bands is
\begin{equation}\label{eqn:Ecv}
	\begin{aligned}
		&E^c_{1/2}=\frac{\varepsilon_k^c+\varepsilon_k^v+\omega_{\text{P}}\pm\sqrt{\left(\varepsilon_k^c-\varepsilon_k^v-\omega_{\text{P}}\right)^2+4|\Delta_k|^2}}{2}
		,\\
		&E^v_{1/2}=\frac{\varepsilon_k^v+\varepsilon_k^c-\omega_{\text{P}}\pm\sqrt{\left(\varepsilon_k^v-\varepsilon_k^c+\omega_{\text{P}}\right)^2+4|\Delta_k|^2}}{2}
	\end{aligned}
\end{equation}
where $E^c_{1/2}$ are the two bands from the conduction band $\varepsilon^c$ and the first Floquet replica ($v^{+1}$) of the valence band $\varepsilon^v+\omega_{\text{P}}$ after hybridizing with the matrix element $\Delta_k$. Since $\Delta_k$ becomes  large  as the driving frequency $\omega_{\text{P}}$ is tuned close to $\omega_\lambda$, the excitonic effect   leads to dramatic corrections to the energy compared to the case without electron-hole interaction, as schematically shown in \fig{fig1}(a).  According to \equa{eqn:delta_k}, the energy distortion as a function of momentum also reflects the electron-hole bound-state wave function $\phi_\lambda(k)$.
Similar effects occur for $E^v_{1/2}$, the hybridized bands from $\varepsilon_k^v$ and $\varepsilon_k^c-\omega_{\text{P}}$ through $\Delta_k$.
Note that as $\omega_{\text{P}}$ approaches  $\omega_\lambda$, 
the vertex in
\equa{eqn:delta_k} appears divergent if there is no excitonic damping.
In fact, the large oscillation amplitude of the excitonic mode must cause a blue shift of the exciton energy via the exciton-exciton interaction~\cite{Wu2015, Sun.2024_dynamical}, so that the divergence is cured.
This effect is contained in the pump-induced corrections to the electron-photon vertex, which we address in future research.

The occupation information of  the Floquet electronic bands could be found from the lesser Green function  $G_{\alpha,\beta}^<(t_1,t_2)= i \langle \alpha^\dagger(t_2) \beta(t_1) \rangle =(G^K+G^A-G^R)/2$, where $\alpha,\beta$ takes values in $\{c,v\}$. For the $v^{+1}$ pole shown in Fig.\,\ref{fig1}(a), we obtain
\begin{equation}\label{eqn:Gv1}
	\begin{aligned}
		G^<_{ v^{+1}}(\omega)=2\pi\mathrm{i}\frac{|\Delta_k|^2}{\left(\varepsilon_k^c-\varepsilon_k^v-\omega_{\text{P}}\right)^2+4|\Delta_k|^2}
		\delta(\omega-E_2^c),
	\end{aligned}
\end{equation}
see SM, Sec.~IV~\cite{Supp}.
It is obvious that the occupation grows with  $\Delta_k$, i.e., the driving strength and the magnitude of the exciton wave function. 

\emph{Phosphorene and MoS$_2$.} To make quantitative predictions for experiments, we present,  in \fig{fig2}, the numerical results of phosphorene~\cite{Wang2015,Tian.2020, Yoon.2021, Wang2021, Carre2021BPExcitons, Lei2023} and monolayer MoS$_2$ using the two-band model in \equa{eq:model} with the parameters from their relevant bands. 
The energy dispersion of the Floquet bands is obtained by the poles of the Green function from solving Eqs.~\eqref{eqn:dyson}, \eqref{eqn:self_energy}, \eqref{eqn:vertex}, and \eqref{eq:wannier}.

For the excitons, we use the two-dimensional Coulomb attraction with the kernel $V_q=2\pi/(\epsilon q)$ in \equa{eq:wannier}, where $\epsilon$ is the dielectric  constant of the three-dimensional environment. It yields a hydrogenic series of bound states~\cite{Yang1991}.
For example, the 1$s$ exciton has the wave function 
$
	\phi_{1s}(k)=\sqrt{8\pi} q_0^2/(q_0^2+k^2)^{3/2}
$
and the binding energy 
$
E_{\text{b}}=me^4/(\hbar^2 \epsilon^2)
$,
 where $q_0=2/a_0$, $a_0=2\epsilon \hbar^2/me^2$ is the exciton Bohr radius and $m=2/(1/m_{{c}}+1/m_{{v}})$ is the reduced mass. 
We note that while two-dimensional screening from the monolayer itself changes the Coulomb attraction  kernel  to the Rytova-Keldysh potential~\cite{Keldysh1979}, it does not bring qualitative changes to the picture~\cite{Shan.2014,Heinz.2014,Xu.2023}; see the SM of Ref.~\cite{Sun.2024_dynamical}.
 
A typical experiment to measure these Floquet bands is Tr-ARPES that probes the photoemission intensity~\cite{Bao2024},
\begin{align}\label{eqn:arpes}
		I_{k}(\varepsilon_k^f)=-\mathrm{i} |A_b|^2 \sum_{\alpha, \beta}M_{f \alpha}(k)M_{\beta f}(k)  G^<_{\alpha \beta}(\varepsilon_k^f-\omega_b)
\,,
\end{align}	
see SM Sec.~VI~\cite{Supp}. 
Here the vector potential of the probe field is ${A}_b(t)={A}_b\mathrm{e}^{-\mathrm{i}\omega_b t}+\text{c.c.}$, 
$M_{fc}$ ($M_{fv}$) is the optical matrix element between free electrons $f$ and the electrons in conduction (valance) band, $\varepsilon_k^f$ is the kinetic energy of the emitted free electron, and $G^<$  is the  lesser Green function from \equa{eqn:dyson} that contains the effect of the interaction and the pumping field. 

In \fig{fig2}(a)-\ref{fig2}(c), 
we show the electronic energy dispersion of  phosphorene under the pump light at three different frequencies.
The dashed lines are the dispersion of the Floquet bands from free theory (without the $H_{\text{ee}}$),
while the color scale shows the Tr-ARPES intensity $I_{k}(\omega+\omega_b)$ from \equa{eqn:arpes} using the excitonic corrected Green functions.
Compared to the free theory,  the excitonic correction leads to dramatic 
downward bending of the $v^{+1}$ band  [e.g., in the region enclosed by the red dotted rectangle in \fig{fig2}(b)] and similar distortion for other bands.
The shape of bending is \emph{not} from a simple band crossing  between the bare $c^0$ and $v^{+1}$ bands since they haven't crossed each other before being hybridized by the optical vertex $\Delta_k$ in \equa{eqn:delta_k}.
Instead, it comes from the strong momentum dependence of the exciton wave function $\phi_{1s}(k)$ within the momentum range set by $q_0$, the size of the exciton wave function, which is understood analytically from  \equa{eqn:Ecv}. 
In the same momentum range, the photoemission intensity of the $v^{+1}$ and $c^0$ bands is also greatly enhanced,  reflecting the exciton-enhanced occupation number predicted by \equa{eqn:Gv1}.
In \fig{fig2}(c), we tune the pump frequency close to the 2$s$ exciton  so that the shape of bending highlights the exciton wave function $\phi_{2s}(k)$ with a momentum range of about $q_0/3$.

The $v^{+1}$ band could also be understood as the ARPES signal of the virtual  1$s$ excitons excited by the pump.
However, we emphasize that the $1s$ exciton is only \emph{virtually} excited by the pump light, in the same sense as a  harmonic oscillator driven  off resonance, different from the situation studied before~\cite{Perfetto_PRB_2016,Rustagi_PRB_2018,Christiansen_PRB_2019, Perfetto.2019, Chan_Giant_PNAS_2023}. 
Resonant excitation of the excitons is not 	required, and the excitonic correction to the electronic Floquet spectra exists at every driving frequency. 
For instance, if the driving frequency $\omega_{\text{P}}$ is shifted, the $v^{+1}$ band, stemming from the replica of the valence band, shifts with it, as clearly shown by \equa{eqn:Ecv}.
Therefore, our approach provides a unified analytical understanding of previous numerical results~\cite{Perfetto_PRB_2016,Rustagi_PRB_2018,Christiansen_PRB_2019, Perfetto.2019, Chan_Giant_PNAS_2023,Park_2024}. We note that  exciton-Floquet effects under resonant pumping have been observed in a recent experiment~\cite{Pareek_Driving_arXiv_2024}, 
while the nontrivial excitonic corrections under off-resonant pump remain to be discovered.
To observe the sharp features in \fig{fig2} using Tr-ARPES~\cite{Mori2023, Pareek_Driving_arXiv_2024} or ultrafast optical technique~\cite{sie_2015_valleyselective,Shan:2021aa,Kobayashi2023}, 
the combined  energy uncertainty from the pump, probe, and the excitonic linewidth~\cite{Wang2015,Tian.2020, Yoon.2021, Wang2021, Lei2023} needs to be within $15 \unit{meV}$.

\emph{Optically active phonons.} We now discuss a rather simple example of collective mode correction to the electronic Floquet spectra in materials driven by light: the case of optically active phonons (often called ``infrared phonons'') coupled to electrons \cite{Hubener2018,BiontaPRR2021,OferPNAS2022,ZhangNP2024,ZhangnpjCM2024}. 
For the  two-band semiconductor described by \equa{eq:model}, the additional Hamiltonian describing the infrared phonon and its coupling to electrons and to light reads
\begin{align}\label{eq:p}
	H_{\text{ep}}= 
	& \omega_0 a^\dagger a
+D\left(a+a^\dagger_{}\right)E(t)
+	
\notag \\
&
\left(a+a^\dagger\right)
\sum_k	\left(
\begin{array}{c}
	{c}_{k}	\\
	{v}_{k}\\
\end{array}
\right)^\dagger \left(
\begin{array}{cc}
	B^{cc}_k & B^{cv}_k\\
	B^{vc}_k & B^{vv}_k \\
\end{array}
\right) \left(
\begin{array}{c}
	{c}_{k}	\\
	{v}_{k}\\
\end{array}
\right) 
\,
\end{align}
where  $a$ ($a^\dagger$)  is the  annihilation (creation) operator for the zero-momentum infrared phonon with the intrinsic frequency $\omega_0$,
$B_k$ is the electron-phonon coupling matrix element,  and $D$ is the linear coupling matrix element between the phonon and the electric field. 

The phonon  is driven linearly by the pump light,  thus modifying the electronic spectra  by contributing a vertex as  another correction to the vertex in \fig{fig1}(c). For instance, its interband component reads
\begin{align}\label{eqn:phonon_vertex}
\Gamma^{cv}_{\text{p}}(\omega_{\text{P}})=
E(\omega_{\text{P}}) 
B_k^{cv} D
\frac{2\omega_0}{\omega_{\text{P}}^2-\omega_0^2+\mathrm{i}\gamma \omega_{\text{P}}}
\end{align}
where the last term comes from the retarded response function of the phonon and we have added a phenomenological damping rate $\gamma$ to it. 
For a single-color drive, \equa{eqn:phonon_vertex} should be simply added to \equa{eqn:vertex} to yield the total vertex that corrects the band in \equa{eqn:Ecv}. Apparently, when the pump frequency is close to the intrinsic phonon frequency, the driven phonon oscillation is resonantly enhanced and so is the phonon-mediated correction to the electronic band.
We note that the existence of an optically active phonon or exciton leads to a nontrivial dielectric function $\epsilon(\omega)$ of the solid that has resonant structures close to their frequencies. The electric field in Eqs.~\eqref{eq:model} and \eqref{eq:p} should be understood not as the external electric field, but as the total electric field after the screening effect, which makes a big difference in three-dimensional materials.

\emph{Discussion.} Before our conclusion, we discuss the logic behind the re-summation  in the driving field $E_0$ for the periodically driven case where ${E}(t)={E}_0(\mathrm{e}^{-\mathrm{i}\omega_{\text{P}} t}+\text{c.c.})$.
The Taylor expansion $G=\sum_n g_{2n} E_0^{2n}$ of the Green function in $E_0$  is re-summed by the Dyson equation in \equa{eqn:dyson} so that one just needs to compute the self-energy, $\Sigma=\sum_n c_{2n} E^{2n}$. This is a common practice which assumes that among all the  contributions to the coefficient $g_{2n}$,  the 
 irreducible diagrams are less important. 
 The general argument for it is that in $g_{2n}$, the irreducible diagrams involve electron propagators at higher frequencies ($\omega+n \omega_{\text{P}}$), which are small if one concerns the low-energy behavior of the Green function. 
Additionally, consider $g_4$ for the excitonic effects as an example; 
one is concerned with the case when $\omega_{\text{P}}$ is close to an exciton resonance so that the vertex $\Gamma^{cv}(\omega_{\text{P}})=\Gamma^{vc}(-\omega_{\text{P}})
\sim 1/(\omega_{\text{P}}-\omega_{\lambda})$ in \equa{eqn:vertex} is large.
Compared to the reducible diagram $G_0 \Sigma G_0 \Sigma G_0$  included by the Dyson's equation with the leading-order self-energy, the second diagram in \fig{fig1}(b) is smaller by the ratio
$a_1 \sim \Gamma^{cv}(-\omega_{\text{P}})
 G_0(\omega+2\omega_{\text{P}})
 \Gamma^{vc}(\omega_{\text{P}})
/
\left[
\Gamma^{cv}(\omega_{\text{P}})
G_0(\omega)
 \Gamma^{vc}(-\omega_{\text{P}})
\right]
\sim (\omega_{\text{P}}-\omega_{\lambda})^2/(\omega_{\text{P}}-E_{\text{g}})^2
\ll 1
$,
which is indeed a small parameter; see SM, Sec.~III~\cite{Supp}.
Nevertheless, we note that for closed ergodic systems, the perturbative result in the pump field is valid only in the prethermal stage before the system is heated to infinite temperature.

With this perturbative approach, more interesting interaction effects are to be discovered for the Floquet electronic spectra, such as polaron effects and effects of collective modes in systems with charge/spin/excitonic orders.
Given improved energy resolution in the pump and probe,  we expect these intriguing phenomena to be observed by Tr-ARPES \cite{Mori2023, Pareek_Driving_arXiv_2024} and ultrafast optical experiments \cite{sie_2015_valleyselective,Shan:2021aa,Kobayashi2023} in the near future.

\emph{Acknowledgments.}
	This work is supported by  the National Key	Research and Development Program of China (Grant No. 2022YFA1204700), the National Natural Science Foundation of China (Grants No. 12374291 and No. 12421004), and the startup grant from Tsinghua University. 
	We thank S. Zhou, J. Li, and X. Yang for helpful discussions.
	
\bibliography{references/Nonequilibrium.bib,references/floquet_references.bib,references/Textbooks.bib,references/Excitons.bib}

\end{document}

%%%%%%%%% Merge with supplemental materials %%%%%%%%%%
%%%%%%%%% Prefix a "S" to all equations, figures, tables and reset the counter %%%%%%%%%%
\setcounter{equation}{0}
\setcounter{figure}{0}
\setcounter{table}{0}
\setcounter{page}{1}
\makeatletter
\renewcommand{\theequation}{S\arabic{equation}}
\renewcommand{\thefigure}{S\arabic{figure}}
\renewcommand{\bibnumfmt}[1]{[S#1]}
\renewcommand{\citenumfont}[1]{S#1}

\title{Supplemental Material for ``Interaction Effects on the Electronic Floquet Spectra: Excitonic Effects''}

\author{Teng Xiao}
\affiliation{State Key Laboratory of Low-Dimensional Quantum Physics and Department of Physics, Tsinghua
	University, Beijing 100084, P. R. China}

\author{Tsan Huang}
\affiliation{State Key Laboratory of Low-Dimensional Quantum Physics and Department of Physics, Tsinghua
	University, Beijing 100084, P. R. China}

\author{Changhua Bao}
\affiliation{State Key Laboratory of Low-Dimensional Quantum Physics and Department of Physics, Tsinghua
	University, Beijing 100084, P. R. China}

%\email{zysun@tsinghua.edu.cn}
\author{Zhiyuan Sun}
\affiliation{State Key Laboratory of Low-Dimensional Quantum Physics and Department of Physics, Tsinghua
	University, Beijing 100084, P. R. China}
\affiliation{Frontier Science Center for Quantum Information, Beijing 100084, P. R. China}

\date{\today}
\maketitle
\tableofcontents

\section{Non-equilibrium Green functions and the Dyson equation}\label{APP:keldysh}
Following Refs.~\cite{Keldysh.1964, Larkin_Ovchinnikov_1975, Kamenev.book}, 
the fermion non-equilibrium   Green functions are defined as
\begin{equation}\label{eqn:Green_path_integral_SI}
\hat{G}_{a b}\left(t, t^{\prime}\right)	
	=
-\mathrm{i}\left\langle\psi_a(t) \bar{\psi}_b\left(t^{\prime}\right)\right\rangle
	=\left(\begin{array}{cc}
		G^{{R}}\left(t, t^{\prime}\right) & G^{{K}}\left(t, t^{\prime}\right) \\
		0 & G^{{A}}\left(t, t^{\prime}\right)
	\end{array}\right),
\end{equation}
where $a/b=1/2$, $
\psi_{1/2}(t)=\frac{1}{\sqrt{2}}\left(\psi^{+}(t)\pm\psi^{-}(t)\right)$,
$
\bar{\psi}_{1/2}(t)=\frac{1}{\sqrt{2}}\left(\bar{\psi}^{+}(t)\mp\bar{\psi}^{-}(t)\right)$,  $\psi^{+}/\psi^{-}$ is the field on the forward/backward part of the closed time contour, 
and $\langle \rangle$ means the path integral average so that the time-ordering symbol is not needed.
For convenience of the readers, we also list the definition of the real-time Green functions from \equa{eqn:Green_path_integral_SI} on the single time axis using the canonical formalism~\cite{Rammer.book2007}: 
\begin{align}
		G^{R}\left( t, t^{\prime}\right)  &=-\mathrm{i} \theta\left(t-t^{\prime}\right)\left\langle\left[\psi(t), \psi^{\dagger}\left(t^{\prime}\right)\right]_{\zeta}\right\rangle 
		=\theta\left(t-t^{\prime}\right)\left(G^{>}\left(t, t^{\prime}\right)-G^{<}\left(t, t^{\prime}\right)\right)
,\notag\\
		G^{A}\left( t, t^{\prime}\right)  &=\mathrm{i} \theta\left(t-t^{\prime}\right)\left\langle\left[\psi(t), \psi^{\dagger}\left(t^{\prime}\right)\right]_{\zeta}\right\rangle 
=-\theta\left(t-t^{\prime}\right)\left(G^{>}\left(t, t^{\prime}\right)-G^{<}\left(t, t^{\prime}\right)\right)	
,\notag\\
	G^{K}\left( t, t^{\prime}\right)  &=-\mathrm{i} \left\langle\left[\psi(t), \psi^{\dagger}\left(t^{\prime}\right)\right]_{-\zeta}\right\rangle 
=G^{>}\left(t, t^{\prime}\right)+G^{<}\left(t, t^{\prime}\right)
\end{align}
where $\langle ... \rangle$ now means $\mathrm{Tr}(\hat{\rho}...)$,  $G^{>}\left(t, t^{\prime}\right)=-\mathrm{i}\left\langle\psi(t) \psi^{\dagger}\left(t^{\prime}\right)\right\rangle$, $G^{<}\left(t, t^{\prime}\right)=-\zeta\mathrm{i}\left\langle\psi^{\dagger}\left(t^{\prime}\right)\psi(t) \right\rangle$, and $\zeta=1/-1$ for bosons/fermions.

The Dyson equation for the non-equilibrium  Green function is
\begin{equation}\label{eqn:dyson_SI}
	\begin{aligned}
		\hat{G}=\hat{G}_0+\hat{G}_0 \hat{\Sigma} \hat{G}, 
\quad
		\hat{G}=\left(\begin{array}{cc}
			G^R & G^K \\
			0 & G^A
		\end{array}\right),
\quad
		\hat{\Sigma} =\left(\begin{array}{cc}
			\Sigma^R & \Sigma^K \\
			0 & \Sigma^A
		\end{array}\right)
	\end{aligned}
\,
\end{equation}
where the products imply matrix multiplication and convolution in their time indices: $(\hat{\Sigma} \hat{G})(t_1,t_3)=\int dt_2 \hat{\Sigma}(t_1, t_2) \hat{G}(t_2, t_3)$.
This is a natural result of re-summation as one performs perturbative expansion in the language of path integral on the Keldysh time contour~\cite{Kamenev.book, Altland.2010}. Its solution in the matrix form is:
\begin{equation}\label{eqn:dyson_solution_SI}
	\begin{aligned}
		\hat{G} & =\left(\hat{G}_0^{-1}-\hat{\Sigma}\right)^{-1} 
		=\left(\begin{array}{cc}
			\left[G_0^{R}\right]^{-1}-\Sigma^R & \left[G_0^{-1}\right]^{K}-\Sigma^K \\
			0 & \left[G_0^{A}\right]^{-1}-\Sigma^A
		\end{array}\right)^{-1} .
	\end{aligned}
\end{equation}
In terms of its retarded, advanced and Keldysh components, \equa{eqn:dyson_SI} reads
\begin{equation}
	\begin{aligned}
		& \begin{cases}G^{R(A)}=G_0^{R(A)}+G_0^{R(A)} \Sigma^{R(A)}G^{R(A)} \\
			G^K=G_0^K+G_0^R \Sigma^R G^K+G_0^R \Sigma^K G^A+G_0^K \Sigma^A G^A\end{cases} \\
	\end{aligned}
\end{equation}
and \equa{eqn:dyson_solution_SI} reads~\cite{Haug_Jauho_2008}
\begin{equation}\label{eqn:Dyson_explicite}
	\begin{aligned}
		&\left\{\begin{array}{ll}
			G^{R(A)}&=\left(1- G_0^{R(A)}\Sigma^{R(A)}\right)^{-1}G_0^{R(A)}=\left(\left[G_0^{{R(A)}}\right]^{-1}-\Sigma^{R(A)}\right)^{-1}\\
			G^K&=G^R\left(G_0^{R}\right)^{-1} G_0^K\left(G_0^{A}\right)^{-1}G^A +G^R\Sigma^K G^A =\left(1+G^R\Sigma^R\right) G_0^K\left(1+\Sigma^AG^A\right) +G^R\Sigma^K G^A
		\end{array}\right. .
	\end{aligned}
\end{equation}
For the two band model, the bare Green functions in the band basis are
\begin{equation}
G_{0,k}^{R/A}(\omega)= \left(\begin{array}{cc}
	\frac{1}{\omega-\varepsilon_k^c \pm i\eta} & 0 \\
	0 & \frac{1}{\omega-\varepsilon_k^v \pm i\eta } 
\end{array}\right)
,\quad
G_{0,k}^K(\omega)= -2\pi i \left[1-2n_F(\omega)\right] \left(\begin{array}{cc}
	\delta(\omega-\varepsilon_k^c) & 0 \\
	0 & \delta(\omega-\varepsilon_k^v) 
\end{array}\right)
\end{equation}
where $n_F(\omega)=1/(e^{(\omega-\mu)/T}+1)$ is the equilibrium Fermi occupation function at temperature $T$. 

\section{Excitonic correction to the electron-photon vertex}\label{APP:vertex}
The first diagram in  Fig.\,1(b) of the main text corresponds to the second order term of $A$ in the self energy:
\begin{equation}
	\begin{aligned}
		\hat{\Sigma}_k(\omega)  =&\hat{\Gamma}_k(\omega, \omega+\omega_{\text{P}})\hat{G}_{0,k}(\omega+\omega_{\text{P}})\hat{\Gamma}_k(\omega+\omega_{\text{P}},\omega)+
		\left(\omega_{\text{P}}\rightarrow -\omega_{\text{P}}\right)
	\end{aligned}
\end{equation}
where $\hat{\Gamma}_k$ is the light-matter coupling vertex, which may be re-normalized by electron-hole interactions. 
We next compute the vertex.
As an example, the self-consistent equation for the interband vertex $\hat{\Gamma}^{cv}_k(\omega+\omega_{\text{P}},\omega)$ shown in Fig.\,1(c) of the main text  is
\begin{equation}\label{eq:gamma}
	\begin{aligned}
		\hat{\Gamma}^{cv}_k(\omega+\omega_{\text{P}},\omega) &=A_0 M_k \hat{\gamma}_c +\frac{\mathrm{i}}{2}
\left[	
		\sum_{\omega^{\prime}, k^{\prime}} V_{k-k^{\prime}} 
		\hat{\gamma}_c
		\hat{G}^{cc}_{0,k^{\prime}}(\omega^{\prime}+\omega_{\text{P}})   \Gamma^{cv}_{k^{\prime}}(\omega^\prime+\omega_{\text{P}},\omega^\prime)
		\hat{G}^{vv}_{0,k^{\prime}}(\omega^{\prime})  
		\hat{\gamma}_q
		+ (\hat{\gamma}_c \leftrightarrow \hat{\gamma}_q)
\right]
	\end{aligned}
\end{equation}
where
$\hat{\gamma}_c=\left(\begin{array}{cc}
	1 & 0 \\
	0 & 1
\end{array}\right)$ 
and 
$\hat{\gamma}_q=\left(\begin{array}{cc}
	0 & 1 \\
	1 & 0
\end{array}\right)$ 
are the `classical' and `quantum' matrices in the Keldysh space~\cite{Kamenev.book}. From the right hand side of  \equa{eq:gamma}, one observes that the vertex   is independent on $\omega$ so that  $\hat{\Gamma}^{cv}_k(\omega+\omega_{\text{P}},\omega)=\hat{\Gamma}^{cv}_k(\omega_{\text{P}})$, i.e., it depends on the photon frequency only.
Taking the trial solution $\hat{\Gamma}^{cv}_k(\omega_{\text{P}})= \Gamma^{cv}_k(\omega_{\text{P}}) \hat{\gamma}_c$, \equa{eq:gamma} is simplified to
\begin{equation}\label{eq:gamma2}
	\begin{aligned}
		\Gamma^{cv}_k(\omega_{\text{P}})&=A_0 M_k-\sum_{ k^{\prime}} V_{k-k^{\prime}} \frac{1}{\omega_{\text{P}}+\varepsilon^v_{k^{\prime}}-\varepsilon^c_{k^{\prime}}+\mathrm{i}0^+} \Gamma^{cv}_{k^{\prime}}(\omega_{\text{P}}),
	\end{aligned}
\end{equation}
a typical Bethe–Salpeter equation.
Before solving \equa{eq:gamma2}, we define the  eigen state problem between an electron and a hole:
\begin{equation}\label{eq:wannier_SI}
	\left(\varepsilon^c_k-\varepsilon^v_k-\omega_\lambda\right) \phi_\lambda(k)=\sum_{k^{\prime}} V_{k-k^{\prime}} \phi_\lambda(k^{\prime})
\end{equation}
where $\omega_\lambda$ and $\phi_\lambda(k)$ are the eigen energy and relative  wave function of the two body problem~\cite{Masaki_1966,Yang1991}. 
The eigen states are classified into bound states ($\omega_\lambda <E_{\text{g}}$) meaning excitons,
and scattering states ($\omega_\lambda > E_{\text{g}}$).
Since $\phi_\lambda(k)$ forms a complete basis of functions, one may expand the vertex as  
$ \Gamma^{cv}_k/\left({\omega+\varepsilon^v_k-\varepsilon^c_k+\mathrm{i}0^+}\right)=\sum_{\lambda}B_{\lambda}\phi_\lambda(k)$, which transforms \equa{eq:gamma2} into
\begin{align}\label{eq:B}
	& \sum_{\lambda}B_{\lambda}\phi_\lambda(k)\left({\omega+\varepsilon^v_k-\varepsilon^c_k+\mathrm{i}0^+}\right)=A_0 M_k-\sum_{ k^{\prime},\lambda} V_{k-k^{\prime}}B_{\lambda}\phi_\lambda(k^{\prime})
\end{align}
Plugging \equa{eq:wannier_SI} into \equa{eq:B} and taking into account the orthogonality properties of $\phi_\lambda$  yields the expansion coefficients:
\begin{align} B_{\lambda}=\frac{A_0}{{\omega_{\text{P}}-\omega_\lambda+\mathrm{i}0^+}}\sum_{k}M_k\phi^*_\lambda(k)
\,.
\end{align}
Finally, one obtains the vertex as an expansion of all electron-hole two-body eigen states:
\begin{equation} \label{eq:vertex_solution}
\Gamma^{cv}_k(\omega_{\text{P}})
=A_0 \sum_{\lambda}
\left(\sum_{k^{\prime}}M_{k^{\prime}}\phi^*_\lambda(k^{\prime})\right)
\frac{\omega_{\text{P}}+\varepsilon^v_k-\varepsilon^c_k}{\omega_{\text{P}}-\omega_\lambda+\mathrm{i}0^+}
\phi_\lambda(k).
\end{equation}
Note that $\phi_\lambda$ include both bound states (excitons) and scattering states.
If an exciton has a finite lifetime, one may change $0^+$ in \equa{eq:vertex_solution} to a phenomenological damping rate $\gamma_{\lambda}$. One observes that the vertex is resonantly enhanced if $\omega_{\text{P}}$ is close to a exciton resonance.

Noticing that  $\left[\Gamma_k(\omega_{\text{P}})\right]^{\dagger}=\Gamma_k(-\omega_{\text{P}})$, the full vertex matrix reads
\begin{equation}\label{Gamma1}
	\begin{aligned}
		\Gamma_k(\omega_{\text{P}})=\left(\begin{array}{cc}
			\Gamma_k^{cc} & \Gamma^{cv}_k(\omega_{\text{P}}) \\
			\left[\Gamma^{cv}_k(-\omega_{\text{P}})\right]^* & \Gamma_k^{vv} 
		\end{array}\right)
	\end{aligned}
,\quad
	\begin{aligned}
		\Gamma_k(-\omega_{\text{P}})=\left(\begin{array}{cc}
			\Gamma_k^{cc} & \Gamma^{cv}_k(-\omega_{\text{P}}) \\
			\left[\Gamma^{cv}_k(\omega_{\text{P}})\right]^* & \Gamma_k^{vv} 
		\end{array}\right),
	\end{aligned}
\end{equation}
where $\Gamma_k^{cc} =A_0 k/m_c$ and $\Gamma_k^{vv} =-A_0 k/m_v$.

\subsection{Contribution from the scattering states}\label{APP:scatter}
In the numerical results shown in the main text, we focus on the simple case of a constant interband optical matrix element $M_k=M$ so that the exciton-corrected vertex is simplified to
\begin{align} \label{eq:vertex_simple}
	\Gamma^{cv}_k(\omega_{\text{P}})
	&=A_0 M (\omega_{\text{P}}+\varepsilon^v_k-\varepsilon^c_k)
	\sum_{\lambda}
	\frac{	\phi^*_\lambda(r=0)\phi_\lambda(k)}
	{\omega_{\text{P}}-\omega_\lambda+\mathrm{i}\gamma_{\lambda}}
	\notag\\
	&=
	A_0 M (\omega_{\text{P}}+\varepsilon^v_k-\varepsilon^c_k)
	\left(
	\sum_{\lambda \in {\text{bound}}}\frac{\phi^*_\lambda(r=0)\phi_\lambda(k)}{\omega_{\text{P}}-\omega_\lambda+\mathrm{i}\gamma_{\lambda}}
	+
	\sum_{\lambda \in {\text{scatter}}}\frac{\phi^*_\lambda(r=0)\phi_\lambda(k)}{\omega_{\text{P}}-\omega_\lambda+\mathrm{i}\gamma_{\lambda}}
	\right)
	\notag\\
	&= 	\Gamma^{cv}_{k, \text{bound}}(\omega_{\text{P}})+\Gamma^{cv}_{k, \text{scatter}}(\omega_{\text{P}})
	\,
\end{align}
which contains bound state and scattering state contributions.
The bound state wavefunction and the excitonic energy is known in analytical forms. We perform the summation over $\lambda_{\text{bound}}$ to a cutoff of 10 to get a converged result.

The summation over the scattering states with a continuous index is hard to perform, which we replace by an approximation formula.
Motivated by the free limit, we assume the scattering state wave function with energy 
$\omega_{\lambda_{\text{scatter}}}$
is a linear combination of free electron-hole states with the relative momenta at all directions but with a fixed magnitude $|\vec{k}|= k_E$ satisfying $\omega_{\lambda_{\text{scatter}}}=k_E^2/m+E_{\text{g}}$  where  $2/m=1/m_c+1/m_v$.

From the relation $\langle k|r=0\rangle=1$, we have $\sum_{\lambda}\phi_\lambda(k)\phi^*_\lambda(r=0)=\sum_{\lambda \in {\text{bound}}}\phi_\lambda(k)\phi^*_\lambda(r=0)+\sum_{\lambda \in {\text{scatter}}}\phi_\lambda(k)\phi^*_\lambda(r=0)=1$. Therefore,
assuming $\gamma_\lambda=\gamma$,  the scattering state contribution could be expressed in terms of bound states as
\begin{equation}
	\begin{aligned}
\Gamma^{cv}_{\vec{k}, \text{scatter}}(\omega_{\text{P}})
&\approx
A_0 M (\omega_{\text{P}}+\varepsilon^v_k-\varepsilon^c_k)
\frac{\sum_{\lambda \in {\text{scatter}}}\phi^*_\lambda(r=0)\phi_\lambda(k)}{\omega_{\text{P}}-(k^2/m+E_{\text{g}})+\mathrm{i}\gamma_{\lambda}}
\notag \\
&=
A_0 M (\omega_{\text{P}}+\varepsilon^v_k-\varepsilon^c_k)
\frac{1-\sum_{\lambda \in {\text{bound}}}\phi^*_\lambda(r=0)\phi_\lambda(k)}{\omega_{\text{P}}-(k^2/m+E_{\text{g}})+\mathrm{i}\gamma_{\lambda}}.
	\end{aligned}
\end{equation}

\subsection{Effect of exciton-exciton nonlinear interactions}
\begin{figure*}[htbp]
	\centering \includegraphics[width=0.5 \linewidth]{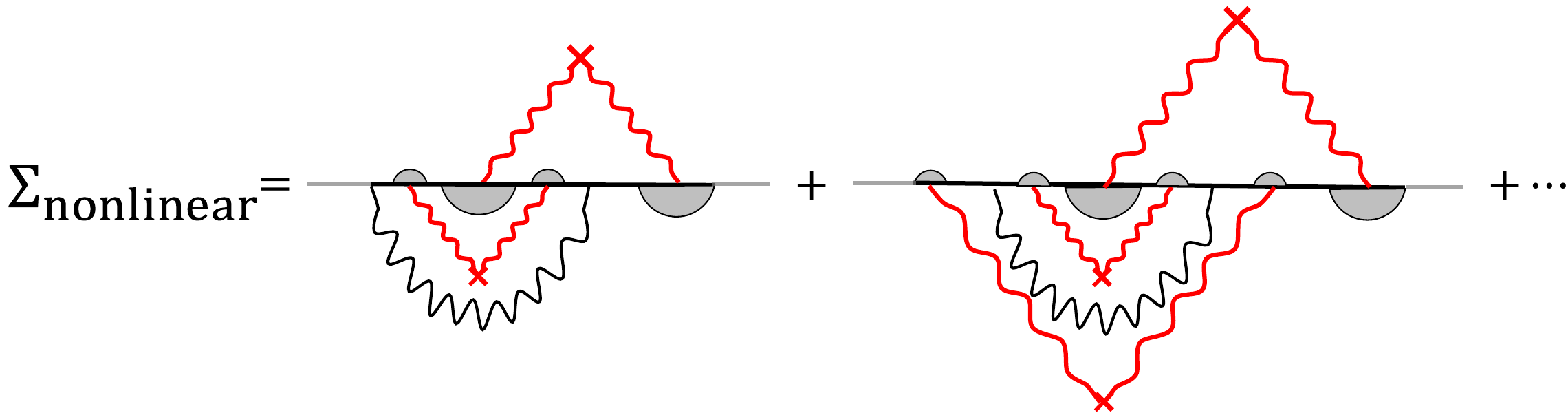} \caption{The self energy diagrams containing the effects of the two-exciton interactions, which lie in the series shown in Fig.~1b in the main text.
	}
	\label{fig:nonlinear_interaction_SI}
\end{figure*}
By making a comparison to the description of exciton-exciton interactions in terms of the Hubbard-Stratonovich field~\cite{Sun.2024_dynamical}, the effect of exciton-exciton interactions are included by the diagrams shown in \fig{fig:nonlinear_interaction_SI}. We leave the evaluation of these diagrams for future research.

\section{Discussion  of the perturbation series}\label{APP:disscusion}
In this section, we discuss the convergence property of the perturbation series for the case of single color pump with no phase randomness.
Same as the main text, we focus on the time averaged Green function $G(\omega)=\int_0^T G(\omega, t_0) dt_0/T$ only.  

\subsection{The Dyson resummation}
The goal of the computation is the Green function.
Dyson re-summation states that one may define the self energy (1PI diagrams defined in the main text) in \equa{eqn:dyson_SI} and perform Taylor expansion of the self energy, which yields the Green function through the Dyson equation.
For the excitonic effects, there is a small parameter that justifies this re-summation when the pump frequency is close to the exciton energy, as discussed in the main text.
For example,
compared to the reducible diagram $G_0 \Sigma G_0 \Sigma G_0$  included by the Dyson's equation with the leading order self energy, the second diagram in Fig.~1b of the main text is smaller by the ratio
\begin{align}
a_1 \sim 
\frac{\Gamma^{cv}(-\omega_{\text{P}})
G_0(\omega+2\omega_{\text{P}})
\Gamma^{vc}(\omega_{\text{P}})
}
{
\Gamma^{cv}(\omega_{\text{P}})
G_0(\omega)
\Gamma^{vc}(-\omega_{\text{P}})
}
\approx\left|\frac{(\omega_{\text{P}}+E_{\text{g}})(\omega_{\text{P}}-\omega_\lambda)}{(\omega_{\text{P}}+\omega_\lambda)(\omega_{\text{P}}-E_{\text{g}})}\right|^2
\ll 1
\,.
\end{align}

\subsection{The Taylor expansion of the self energy}
\begin{figure*}[htbp]
	\centering \includegraphics[width=\linewidth]{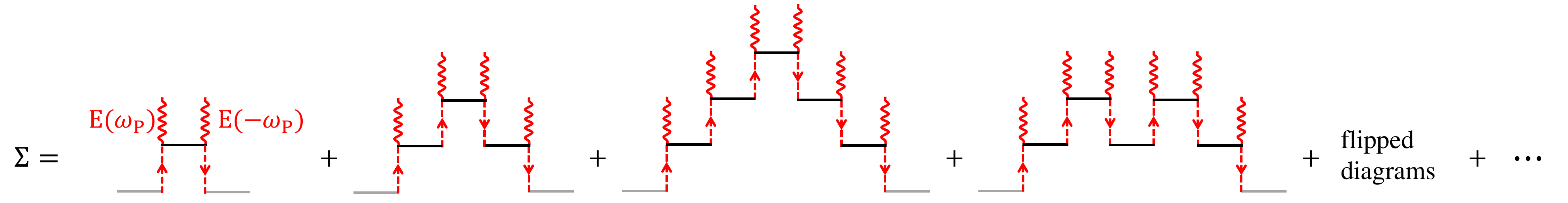} \caption{The  diagrammatic expansion of the self-energy $\Sigma$. The red wavy lines represent the pump light, which are connected to the steps shown by red dashed lines representing the rising and lowering of frequencies by $\omega_{\text{P}}$.
		The gray lines represent external electron legs and should not be included when computing this self energy. 
	The height of the self energy could be defined as the maximum height of the ladder. For example, the  heights of the diagrams are $1,2,3,2$ from left to right in the figure.
	}
	\label{fig:diagrams_SI}
\end{figure*}

The Taylor expansion $\Sigma=\sum_n c_{2n} E^{2n}=\sum_n c^\prime_{2n} a_2^{2n}$ of the self energy itself  could be viewed as an expansion in a dimensionless small parameter proportional to the driving field, which is about $a_2 \sim A_0 M/\omega_{\text{P}}$  for large $n$ as long as $n \omega_{\text{P}}$ is much larger than the relevant electronic band width.
The Taylor expansion series could be grouped diagrammatically in \fig{fig:diagrams_SI}.
At order $E_{\text{P}}^{2n}$ for $n\geq 1$, the number of distinct self energies equals  twice the $n-1$ Catalan number:
$
	N^{2n}_{\text{P}}=\frac{2}{n}\binom{2 n-2}{n-1}
$.

The value of each self energy could be read from the diagram directly. For the convenience of readers not comfortable with diagrams, we also show the symbolic derivations. 	Assuming that $\hat{G}_0(t,t')$ is the equilibrium green function and $\hat{V}(t)$ is the time-periodic pumping term, 
the full Green function may be written as the `grand Dyson equation'
\begin{equation}\label{eqn:grand_Dyson_SI}
	\begin{aligned}
		\hat{G}(t,t')=\hat{G}_0(t,t')+\int\text{d}\tau\hat{G}_0(t,\tau) \hat{V}(\tau) \hat{G}(\tau,t').
	\end{aligned}
\end{equation}
Iterating it, the result at every order of $V$ can be obtained. As in the main text, we consider only  the frequency-conserved Green function as the frequency-non-conserved ones  would vanish after time-average and could hardly be resolved in typical ultra-fast experiments.
Therefore, only even orders of $V$ are needed.
Expanding \equa{eqn:grand_Dyson_SI} to fourth order of $V$ yields
\begin{equation}
	\begin{aligned}
		\hat{G}(\omega)=&\hat{G}_0(\omega)+\hat{G}_0(\omega)\left[\int\text{d}\omega_{\text{P}} \hat{V}(-\omega_{\text{P}}) \hat{G}_0(\omega+\omega_{\text{P}})\hat{V}(\omega_{\text{P}})\right]\hat{G}_0(\omega)+\\
		&\hat{G}_0(\omega)\left\{\left[\int\text{d}\omega_{\text{P}} \hat{V}(-\omega_{\text{P}}) \hat{G}_0(\omega+\omega_{\text{P}})\hat{V}(\omega_{\text{P}})\right]\hat{G}_0(\omega)\left[\int\text{d}\bar{\omega}_P \hat{V}(-\bar{\omega}_P) \hat{G}_0(\omega+\bar{\omega}_P)\hat{V}(\bar{\omega}_P)\right]\right.+\\
		&\left.\left[\int\text{d}\omega_{\text{P}} \hat{V}(-\omega_{\text{P}}) \hat{G}_0(\omega+\omega_{\text{P}})\hat{V}(-\omega_{\text{P}})\hat{G}_0(\omega+2\omega_{\text{P}})\hat{V}(\omega_{\text{P}}) \hat{G}_0(\omega+\omega_{\text{P}})\hat{V}(\omega_{\text{P}})\right]\right\}\hat{G}_0(\omega).
	\end{aligned}
\end{equation}
Comparing to the Dyson's equation for the  equal-frequency Green function in \equa{eqn:dyson_SI}, the self-energy at order $V^2$ (first diagram in \fig{fig:diagrams_SI}) is therefore
\begin{equation}
	\begin{aligned}
		\hat{\Sigma}^{(2)}=\int\text{d}\omega_{\text{P}} \hat{V}(-\omega_{\text{P}}) \hat{G}_0(\omega+\omega_{\text{P}})\hat{V}(\omega_{\text{P}})
	\end{aligned}
\end{equation}
and its flipped version ($\omega_{\text{P}} \rightarrow -\omega_{\text{P}} $).
The self-energy at order $V^4$ (second diagram in \fig{fig:diagrams_SI}) 
\begin{equation}
	\begin{aligned}
		\hat{\Sigma}^{(4)}=
		\int\text{d}\omega_{\text{P}} \hat{V}(-\omega_{\text{P}}) \hat{G}_0(\omega+\omega_{\text{P}})\hat{V}(-\omega_{\text{P}})\hat{G}_0(\omega+2\omega_{\text{P}})\hat{V}(\omega_{\text{P}}) \hat{G}_0(\omega+\omega_{\text{P}})\hat{V}(\omega_{\text{P}})
	\end{aligned}
\end{equation}
and its flipped version.

\section{Excitonic correction to the retarded and lesser Green functions}\label{APP:Green}
The retarded Green function is found from \equa{eqn:Dyson_explicite} with the retarded  self energy
\begin{equation}
\begin{aligned}\label{eqn:sigma_R_SI}
	{\Sigma}^R(\omega) =&{\Gamma}_k(-\omega_{\text{P}}){G}^R_{0,k}(\omega+\omega_{\text{P}}){\Gamma}_k(\omega_{\text{P}})
	+(\omega_{\text{P}} \rightarrow -\omega_{\text{P}}).
\end{aligned}
\end{equation}
being the first diagram in \fig{fig:diagrams_SI} (the lowest order in the pump field).
The retarded  self energy is a two-by-two matrix in the band basis with the matrix elements 
\begin{align}\label{eqn:sigma_R_elements_SI}
		{\Sigma}^R_{cc}=&
			|\Gamma_k^{cc}|^2G_{0,cc}^{R}(\omega+\omega_{\text{P}}) +	|\Gamma_k^{cv}(-\omega_{\text{P}})|^2G_{0,vv}^{R}(\omega+\omega_{\text{P}})
			+(\omega_{\text{P}} \rightarrow -\omega_{\text{P}}),
\notag\\
		{\Sigma}^R_{cv}=&\Gamma_k^{cc}\Gamma_k^{cv}(\omega_{\text{P}})G_{0,cc}^{R}(\omega+\omega_{\text{P}}) +	\Gamma_k^{cv}(-\omega_{\text{P}})\Gamma_k^{vv}G_{0,vv}^{R}(\omega+\omega_{\text{P}})
		+(\omega_{\text{P}} \rightarrow -\omega_{\text{P}}),
\notag\\
			{\Sigma}^R_{vc}=&
	\left[\Gamma_k^{cv}(\omega_{\text{P}})\right]^*\Gamma_k^{cc}G_{0,cc}^{R}(\omega+\omega_{\text{P}}) +	\Gamma_k^{vv}\left[\Gamma_k^{cv}(-\omega_{\text{P}})\right]^*G_{0,vv}^{R}(\omega+\omega_{\text{P}}) 
			+(\omega_{\text{P}} \rightarrow -\omega_{\text{P}}),
\notag\\
		{\Sigma}^R_{vv}=&	|\Gamma_k^{vv}|^2G_{0,vv}^{R}(\omega+\omega_{\text{P}}) +	|\Gamma_k^{cv}(\omega_{\text{P}})|^2G_{0,cc}^{R}(\omega+\omega_{\text{P}})
			+(\omega_{\text{P}} \rightarrow -\omega_{\text{P}}).
\end{align}

The lesser Green function $G^< =(G^K+G^A-G^R)/2$ is found from \equa{eqn:Dyson_explicite} as
\begin{equation}\label{eqn:sigma_lesser_elements_SI}
		\begin{aligned}
			G^<=&G^R\left[G_0^{R}\right]^{-1} G_0^<\left[G_0^{A}\right]^{-1}G^A +G^R\Sigma^< G^A \\
			=&\left(\begin{array}{cc}
				G^R_{cv}\left[G_{0,vv}^{R}\right]^{-1} G_{0,vv}^<\left[G_{0,vv}^{A}\right]^{-1}G^A_{vc}& 	G^R_{cv}\left[G_{0,vv}^{R}\right]^{-1} G_{0,vv}^<\left[G_{0,vv}^{A}\right]^{-1}G^A_{vv} \\
				G^R_{vv}\left[G_{0,vv}^{R}\right]^{-1} G_{0,vv}^<\left[G_{0,vv}^{A}\right]^{-1}G^A_{vc} & 	G^R_{vv}\left[G_{0,vv}^{R}\right]^{-1} G_{0,vv}^<\left[G_{0,vv}^{A}\right]^{-1}G^A_{vv}
			\end{array}\right)+\\
			&\left(\begin{array}{cc}
				G^R_{cc}\Sigma^< _{cc}G^A_{cc}+G^R_{cv}\Sigma^< _{vc}G^A_{cc}+G^R_{cc}\Sigma^< _{cv}G^A_{vc}+G^R_{cv}\Sigma^< _{vv}G^A_{vc}& 	G^R_{cc}\Sigma^< _{cc}G^A_{cv}+G^R_{cv}\Sigma^< _{vc}G^A_{cv}+G^R_{cc}\Sigma^< _{cv}G^A_{vv}+G^R_{cv}\Sigma^< _{vv}G^A_{vv} \\
				G^R_{vc}\Sigma^< _{cc}G^A_{cc}+G^R_{vv}\Sigma^< _{vc}G^A_{cc}+G^R_{vc}\Sigma^< _{cv}G^A_{vc}+G^R_{vv}\Sigma^< _{vv}G^A_{vc} & 	G^R_{vc}\Sigma^< _{cc}G^A_{cv}+G^R_{vv}\Sigma^< _{vc}G^A_{cv}+G^R_{vc}\Sigma^< _{cv}G^A_{vv}+G^R_{vv}\Sigma^< _{vv}G^A_{vv}
			\end{array}\right),
		\end{aligned}
\end{equation}
where
\begin{equation}\label{eqn:sigma_lesser_sigma_SI}
	\begin{aligned}
		{\Sigma}^<=&\left(\begin{array}{cc}
			|\Gamma_k^{cv}(-\omega_{\text{P}})|^2G_{0,vv}^{<}(\omega+\omega_{\text{P}})&	\Gamma_k^{cv}(-\omega_{\text{P}})\Gamma_k^{vv}G_{0,vv}^{<}(\omega+\omega_{\text{P}}) \\
			\Gamma_k^{vv}\left[\Gamma_k^{cv}(-\omega_{\text{P}})\right]^*G_{0,vv}^{<}(\omega+\omega_{\text{P}}) & 	|\Gamma_k^{vv}|^2G_{0,vv}^{<}(\omega+\omega_{\text{P}})
		\end{array}\right)+\\
		&\left(\begin{array}{cc}
			|\Gamma_k^{cv}(\omega_{\text{P}})|^2G_{0,vv}^{<}(\omega-\omega_{\text{P}})& 		\Gamma_k^{cv}(\omega_{\text{P}})\Gamma_k^{vv}G_{0,vv}^{<}(\omega-\omega_{\text{P}}) \\
			\Gamma_k^{vv}\left[\Gamma_k^{cv}(\omega_{\text{P}})\right]^*G_{0,vv}^{<}(\omega-\omega_{\text{P}}) & 	|\Gamma_k^{vv}|^2G_{0,vv}^{<}(\omega-\omega_{\text{P}}) 
		\end{array}\right)
	\end{aligned}
\end{equation}
is the lesser self energy.

\subsection{Close to the exciton resonance}\label{APP:solution}
Close to the exciton resonances, the Green functions can be simplified by neglecting the terms containing $\Gamma_k^{cc}$ and $\Gamma_k^{vv}$ which do not contain the exciton contributions.
In this limit, the self energy in \equa{eqn:sigma_R_SI} has only diagonal components, see \equa{eqn:sigma_R_elements_SI}. If one further neglects the far-from-resonant terms such as $\Gamma_k^{cv}(-\omega_{\text{P}})$, the retarded self energy is simplified to
\begin{equation}
	\begin{aligned}
		{\Sigma}^R
		=|\Delta_k|^2\left(\begin{array}{cc}
			G_{0,vv}^{R}(\omega-\omega_{\text{P}})& 	0 \\
			0 & 	G_{0,cc}^{R}(\omega+\omega_{\text{P}})
		\end{array}\right),
	\end{aligned}
\end{equation}
where $\Delta_k=\Gamma_k^{cv}(\omega_{\text{P}})$ is the exciton-modified vertex in \equa{eq:vertex_solution}.
From \equa{eqn:Dyson_explicite}, the retarded Green functions are found as
\begin{equation}
	\begin{aligned}
		G^R_{cc}	
		 &=
		 \frac{1}
		{\left[G_{0,cc}^{R}(\omega)\right]^{-1}-|\Delta_k|^2 G_{0,vv}^{R}(\omega-\omega_{\text{P}})}
		=\frac{\omega-\omega_{\text{P}}-\varepsilon_k^v+\mathrm{i}0^+}
		{\left(\omega-E_1^c\right)\left(\omega-E_2^c\right)+\mathrm{i}0^+\left(2\omega-E_1^c-E_2^c\right)},
\notag\\
		G^R_{vv}	& 
=\frac{1}
{\left[G_{0,vv}^{R}(\omega)\right]^{-1}-|\Delta_k|^2 G_{0,cc}^{R}(\omega+\omega_{\text{P}})}
 =\frac{\omega+\omega_{\text{P}}-\varepsilon_k^c+\mathrm{i}0^+}
{\left(\omega-E_1^v\right)\left(\omega-E_2^v\right)+\mathrm{i}0^+\left(2\omega-E_1^v-E_2^v\right)},
	\end{aligned}
\end{equation}
where 
\begin{equation}
	\begin{aligned}
		E^c_{1/2}=\frac{\varepsilon_k^c+\varepsilon_k^v+\omega_{\text{P}}\pm\sqrt{\left(\varepsilon_k^c-\varepsilon_k^v-\omega_{\text{P}}\right)^2+4|\Delta_k|^2}}{2}
,\quad
		E^v_{1/2}=\frac{\varepsilon_k^v+\varepsilon_k^c-\omega_{\text{P}}\pm\sqrt{\left(\varepsilon_k^v-\varepsilon_k^c+\omega_{\text{P}}\right)^2+4|\Delta_k|^2}}{2}
	\end{aligned}
\,.
\end{equation}
The advanced Green functions are related to the retarded ones by $G^A=G^{R\ast}$.

The lesser self energy from \equa{eqn:sigma_lesser_sigma_SI}  is
\begin{equation}
	\begin{aligned}
		{\Sigma}^<
		=|\Delta_k|^2\left(\begin{array}{cc}
			G_{0,vv}^{<}(\omega-\omega_{\text{P}})& 	0 \\
			0 & 	0
		\end{array}\right)
	\end{aligned}
\end{equation}
considering the property of the ground state  ($\hat{c}_k\left|0\right\rangle=\hat{v}_k^{\dagger}\left|0\right\rangle=0$).
From \equa{eqn:sigma_lesser_elements_SI}, we obtain the lesser Green functions as
\begin{align}
		G^<_{cc}	& =G^R_{cc}\Sigma^< _{cc}G^A_{cc} 
		 =|\Delta_k|^2\left|G^R_{cc}\right|^2G_{0,vv}^{<}(\omega-\omega_{\text{P}})
 =2\pi\mathrm{i}\frac{|\Delta_k|^2}{\left|E_1^c-E_2^c\right|^2}\left(\delta(\omega-E_1^c)+\delta(\omega-E_2^c)\right)
\notag	\\
		&=2\pi\mathrm{i}\frac{|\Delta_k|^2}{\left(\varepsilon_k^c-\omega_{\text{P}}-\varepsilon_k^v\right)^2+4|\Delta_k|^2}\left(\delta(\omega-E_1^c)+\delta(\omega-E_2^c)\right),
\notag\\
		G^<_{vv}	& =G^R_{vv}\left[G_{0,vv}^{R}\right]^{-1} G_{0,vv}^<\left[G_{0,vv}^{A}\right]^{-1}G^A_{vv} 
		=\left|G^R_{vv}\right|^2\left|\left[G_{0,vv}^{R}\right]^{-1}\right|^2G_{0,vv}^{<}(\omega)
=2\pi\mathrm{i}\frac{\left(\omega+\omega_{\text{P}}-\varepsilon_k^c\right)^2}{\left|E_1^v-E_2^v\right|^2}\left(\delta(\omega-E_1^v)+\delta(\omega-E_2^v)\right)
\notag\\
		&=2\pi\mathrm{i}\frac{\left(\omega+\omega_{\text{P}}-\varepsilon_k^c\right)^2}{\left(\omega_{\text{P}}+\varepsilon_k^v-\varepsilon_k^c\right)^2+4|\Delta_k|^2}\left(\delta(\omega-E_1^v)+\delta(\omega-E_2^v)\right).
\end{align}
We note that the occupation number from these results are for a monochromatic pump turned on sharply  at time zero. If the pump pulse is slowly turned on, a non-perturbative Landau-Zener analysis is required to obtain the correct occupation information.

\section{Connection to the conventional Floquet approach}\label{APP:floquet}
If one neglects the electron-hole interaction $H_{\text{ee}}$ in $H$ of Eq.~3 in the main text, the corresponding Floquet matrix $F$ using the conventional Floquet approach is
\begin{equation}
	\begin{aligned}
		F=\left(
		\begin{array}{lllll}
			\ddots	 \\
			V_f & H_0-\omega_{\text{P}} & V_f & &\\
			& V_f & H_0 & V_f \\
			& & V_f & H_0+\omega_{\text{P}} & V_f \\
			& & & &\ddots
		\end{array}
		\right),	
	\end{aligned}
\end{equation}
where $H_0=\left(
\begin{array}{ll}
	\varepsilon_k^c &  \\
	& \varepsilon_k^v
\end{array}
\right)$,
and 
$V_f=A\left(
\begin{array}{ll}
	k/m_c & M_k \\
	M_k & -k/m_v
\end{array}
\right)$ is the light matter coupling vextex.
With the interaction modified vertex $\Gamma_k$
derived in Sec.~\ref{APP:vertex}, 
one may simply replace the bare coupling $V_f$  by the interaction corrected vertex, so that the
Floquet matrix $F_c$ becomes
\begin{equation}
	\begin{aligned}
		F_c=\left(
		\begin{array}{lllll}
			\ddots	 \\
			\Gamma_k(-\omega_{\text{P}}) & H_0-\omega_{\text{P}} & \Gamma_k(\omega_{\text{P}}) \\
			&\Gamma_k(-\omega_{\text{P}}) & H_0 & \Gamma_k(\omega_{\text{P}}) \\
			&& \Gamma_k(-\omega_{\text{P}}) & H_0+\omega_{\text{P}} & \Gamma_k(\omega_{\text{P}}) \\
			&& & &\ddots
		\end{array}
		\right).
	\end{aligned}
\end{equation}
By diagonalizing this Floquet matrix $F_c$, one obtains its eigenvalues as the quasi-energies, resulting in the same interaction-corrected Floquet bands as those from the Green function method. 
However, the exact correspondence is complicated by a finite frequency cutoff.
 For a cutoff frequency index $\pm \omega_{\text{P}}$ of the Floquet matrix $F_c$, the quasi-energies 
correspond to keeping the $O(E_{\text{P}}^2)$ self-energy used in the main text.
For a cutoff frequency $\pm n\omega_{\text{P}}$, the quasi energies from the Floquet matrix correspond to summing all the self energies whose heights lie within $\pm n$ in
\fig{fig:diagrams_SI}. However, it no longer corresponds to a certain order of the self energy in the pump field.

In the Green function approach, one may define $\hat{\Sigma}^+_{(n)}$  as the sum of all self energy diagrams whose heights are positive but not higher than 
$+n$. 
Similarly, the Green function
$
\hat{G}^+_{(n)}=\left[{\hat{G}_0}^{-1}-\hat{\Sigma}^+_{(n)}\right]^{-1}
$ contains all the 
diagrams whose heights are positive but not higher than 
$+n$. 
They satisfy the recursive relation:
\begin{equation}
	\hat{\Sigma}^+_{(n+1)}(\omega)=\hat{\Gamma}(-\omega_{\text{P}})\hat{G}^+_{(n)}(\omega+\omega_{\text{P}})\hat{\Gamma}(\omega_{\text{P}}).
\end{equation}
Numerically, the self energy could be computed height by height in this way. The  the downward self-energy $\hat{\Sigma}^-_{(-n)}$ may be obtained in a similar way.
The sum $\hat{\Sigma}^+_{(n)}+\hat{\Sigma}^-_{(-n)}$ thus corresponds to all the self energies whose heights lie within $\pm n$.

% order Floquet band  explicitly, we perform the upward self-energy $\hat{\Sigma}^+_{(n\omega_{\text{P}})}$ containing the first $n$ height ($n\omega_{\text{P}},n>0$) from the self-energy expanded in the pump $\hat{\Sigma}^{(2n)}$ (Sec. \ref{APP:disscusion}):

%$\hat{G}^+_{n\omega_{\text{P}}}(\omega)=\hat{G}_0(\omega)+\hat{G}_0(\omega)\hat{\Sigma}^+_{(n\omega_{\text{P}})}\hat{G}^+_{n\omega_{\text{P}}}(\omega)$,
%i.e.

\section{The ARPES  signal}\label{APP:arpes}

The probe part of the Hamiltonian for ARPES is 
\begin{equation}
	\begin{aligned}
		 {H}_f=\sum_{k} \varepsilon_k^{f} {f}_{k}^{\dagger} {f}_{k},
\quad
		{H}_{b}=A_b(t)\sum_{{k}}\left( M_{fc}(k) {f}_{k}^{\dagger} {c}_{k} +M_{fv}(k) {f}_{k}^{\dagger} {v}_{{k}} +\text{h.c.}\right)
	\end{aligned}
\end{equation}
where $\varepsilon_{{k}}^{f}={k^2}/{2 m_e}$ is the kinetic energy of the free electron with $f_k$ being its annihilation operator, ${A}_b(t)={A}_b\mathrm{e}^{-\mathrm{i}\omega_b t}+\text{c.c.}$ is the vector potential of the probe field, $M_{fc}$ ($M_{fv}$) is the optical matrix element between free electrons $f$ and the electrons in conduction (valence) band.
The time-accumulated ARPES intensity is 
\begin{equation}
	\begin{aligned}
		I_k^{\text{sum}}=\int_{t_0}^{t_f} \text{d} t \frac{\partial}{\partial t} \rho_{k}^{f}
		=\int_{t_0}^{t_{f}} \text{d} t I_{k},
	\end{aligned}
\end{equation}
where $I_{k}$ is the ARPES intensity, $\rho_{k}^{f}=\left\langle{f}_{k}^{\dagger} {f}_{k}\right\rangle =-\mathrm{i}G^<_{ff}(t,t)$ and $G^<_{ff}$ is the lesser Green function of the free space electrons collected by the detector.
To eliminate the time derivative, we use the Heisenberg equation of motion
$i \partial_t {A}=[{A}, {H}] $ :
\begin{equation}
	\begin{aligned}
		I_{k}=&\frac{\partial}{\partial t} \rho_{k}^{f}=-\mathrm{i}\frac{\partial}{\partial t}G^<_{ff}(t,t)
		= 2\,\text{Re}\left({\sum_{\alpha=\{c,v\}} M_{\alpha f}(k) A_b(t)G^<_{f\alpha}(t,t)}\right)
		\label{eqn:Ik_Gfm}
	\end{aligned}
\end{equation}
where $G^<_{f\alpha}(t,t^{\prime})=\mathrm{i}\left\langle  {\alpha}^{\dagger}_k(t^{\prime}) {f}_k(t)\right\rangle$.
For the ground state $\left|0\right\rangle$ satisfying ${f}_k\left|0\right\rangle={c}_k\left|0\right\rangle={v}_k^{\dagger}\left|0\right\rangle=0$,  the bare lesser, retarded and advanced Green functions are
\begin{align}
	G_0^{<}(t,t^{\prime}) &= \text{diag}\left(G_{0,ff}^{<}(t,t^{\prime}),\,
	G_{0,cc}^{<}(t,t^{\prime}) ,\,
	G_{0,vv}^{<}(t,t^{\prime})
	\right) 
	= 
	\text{diag}\left(0,\,
	0,\,
	\mathrm{i}\mathrm{e}^{\mathrm{i}\varepsilon_{k}^v(t^{\prime}-t)}
	\right),
	\notag\\
	G_0^{R}(t,t^{\prime}) &= \text{diag}\left( 
	G_{0,ff}^{R}(t,t^{\prime}) ,\,
	G_{0,cc}^{R}(t,t^{\prime}),\,
	G_{0,vv}^{R}(t,t^{\prime})
	\right),
	\notag\\
	G_0^{A}(t,t^{\prime}) &=\text{diag}\left( 
	G_{0,ff}^{A}(t,t^{\prime})  ,\, G_{0,cc}^{a}(t,t^{\prime}) ,\, G_{0,vv}^{A}(t,t^{\prime})
	\right),
\end{align}
where $G_{0,ff/cc/vv}^{R}(t,t^{\prime})=-\theta\left(t-t^{\prime}\right) \mathrm{i}\mathrm{e}^{-\mathrm{i}\varepsilon_{k}^{f/c/v}(t-t^{\prime})}$ and  $G_{0,ff/cc/vv}^{A}(t,t^{\prime})=\theta\left(t^{\prime}-t\right) \mathrm{i}\mathrm{e}^{\mathrm{i}\varepsilon_{k}^{f/c/v}(t^{\prime}-t)}$.
Using the Langreth rules to the grand Dyson equation, we obtain
\begin{equation}\label{eqn:Gfm}
	\begin{aligned}
		G^<_{f\alpha}(t,t')=& \sum_{\beta=\{c,v\}}\int^{+\infty}_{-\infty}\text{d}\bar{t} G^R_{0,ff}(t,\bar{t})M_{f\beta}(k)A_b(\bar{t}) G^<_{\beta\alpha}(\bar{t},t').
	\end{aligned}
\end{equation}
to the first order in the probe field.
From \eqref{eqn:Ik_Gfm},  the ARPES intensity is related to the lesser Green function of electrons inside the material~\cite{Bao2024}:
\begin{equation}\label{eqn:I_Glesser}
	\begin{aligned}
		I_{k}=& 2\,\text{Re}\left({  \sum_{\alpha,\beta}
			M_{\beta f}(k)A_b(t)\int^{+\infty}_{-\infty}\text{d}\bar{t} G^R_{0,ff}(t,\bar{t})M_{f\alpha}(k)A_b(\bar{t}) G^<_{\alpha\beta}(\bar{t},t)}\right)\\
		=&-\mathrm{i} |A_b|^2 \sum_{\alpha,\beta}M_{f\alpha}(k)M_{\beta f}(k)G^<_{\alpha\beta}(\epsilon_k^f\pm\omega_b).
	\end{aligned}
\end{equation}
Here $G^<$  is the corrected Green function that contains the effect of the interaction and the pumping field.
For more realistic cases, the probe field is a pulse ${A}_b(t)={A}_bs_b(t)\mathrm{e}^{-\mathrm{i}\omega_b t}+\text{c.c.}$ where $s_b(t)=\exp(-\frac{(t-t_d)^2}{2\sigma^2})$. 
The time-accumulated ARPES intensity is further broadened to
\begin{equation}
	\begin{aligned}
		I_k^{\text{sum}}=&\int_{t_0}^{t_{f}} \text{d} t I_{k}=-\mathrm{i}2\pi\sigma^2 |A_b|^2 \sum_{\alpha,\beta}M_{f\alpha}(k)M_{\beta f}(k)\int\frac{\text{d}\omega}{2\pi}\mathrm{e}^{-\sigma^2(\epsilon_k^f\pm\omega_b-\omega)^2}G^<_{\alpha\beta}(\omega).
	\end{aligned}
\end{equation}
where we take $t_f\to +\infty$ and $t_0\to -\infty$.  However, we take $\sigma=\infty$ for the numerical plots in the main text, meaning single colored probe light.

\bibliography{Nonequilibrium.bib,floquet_references.bib,Textbooks.bib,Excitons.bib}